\documentclass[twocolumn,twocolappendix]{aastex63}

\usepackage{graphicx}
\usepackage{natbib}
\usepackage{float}
\usepackage{xcolor}
\usepackage[utf8]{inputenc}

\newcommand{\msun}{M$_{\odot}$}

\title{Polstar X1/X2}

\begin{document}

\title{UV Spectropolarimetry with Polstar: Protoplanetary Disks}

\author{John P. Wisniewski}
\affiliation{Department of Physics and Astronomy, University of Oklahoma}
\author{Andrei V. Berdyugin}
\affiliation{Tuorla Observatory, Department of Physics and Astronomy, University of Turku, Finland}
\author{Svetlana V. Berdyugina}
\affiliation{Leibniz-Institut f\"ur Sonnenphysik, Freiburg, Germany}
\author{William C. Danchi}
\affiliation{NASA GSFC}
\author{Ruobing Dong}
\affiliation{Department of Physics and Astronomy, University of Victoria}
\author{Ren\'e D. Oudmaijer}
\affiliation{School of Physics and Astronomy, University of Leeds}
\author{Vladimir S. Airapetian}
\affiliation{NASA GSFC/SEEC}
\author{Sean D. Brittain}
\affiliation{Clemson University}
\author{Ken Gayley}
\affiliation{University of Iowa}
\author{Richard Ignace}
\affiliation{East Tennessee State University}
\author{Maud Langlois}
\affiliation{CNRS, ENS, UCBL, Observatoire de Lyon, France}
\author{Kellen D. Lawson}
\affiliation{Department of Physics and Astronomy, University of Oklahoma}
\author{Jamie R. Lomax}
\affiliation{Physics Department, United States Naval Academy, 572C Holloway Rd, Annapolis, MD 21402, USA}
\author{Motohide Tamura}
\affiliation{NAOJ}
\author{Jorick S. Vink}
\affiliation{Armagh Observatory and Planetarium}
\author{Paul A. Scowen}
\affiliation{NASA GSFC}

\begin{abstract}

Polstar is a proposed NASA MIDEX mission that would feature a high resolution UV spectropolarimeter capable of measure all four Stokes parameters onboard a 60cm telescope. The mission would pioneer the field of time-domain UV spectropolarimetry.  Time domain UV spectropolarimetry offers the best resource to determine the geometry and physical conditions of protoplanetary disks from the stellar surface to $<$5 AU.  We detail two key objectives that a dedicated time domain UV spectropolarimetry survey, such as that enabled by Polstar, could achieve: 1) Test the hypothesis that magneto-accretion operating in young planet-forming disks around lower-mass stars transitions to boundary layer accretion in planet-forming disks around higher mass stars; and 2) Discriminate whether transient events in the innermost regions of planet-forming disks of intermediate mass stars are caused by inner disk mis-alignments or from stellar or disk emissions. \vspace{1cm}

\end{abstract}

\section{Introduction}
\subsection{Herbig AeBe systems}

Herbig AeBe objects are pre-main sequence stars (PMS) of spectral types A or B and of intermediate masses ($2M_\odot\lesssim M_\star\lesssim8M_\odot$) \citep{wat98}. They are still contracting on the radiative track. They are typically embedded in gaseous envelopes and surrounded by circumstellar disks. These objects were first identified as a population by \citet{herbig60}, who found a sample of AB stars with emission lines in the spectrum (in particular H$\alpha$) and associated with reflection nebula. 

One motivation to study Herbig AeBe objects is that they provide a vital link between high and low mass stars in pre-main sequence evolution. Herbig stars are the high mass counterparts of T Tauri objects, which are PMS of lower masses also surrounded by circumstellar disks. As the stellar mass increases, the stellar density in the surrounding environment increases, the pre-main sequence evolutionary timescale shortens, the stellar radiation field becomes stronger, but the stellar magnetic activity decreases. All these affect the dynamics and chemistry in the circumstellar disk, and ultimately, how planets form around stars with different masses.

UX Ori-type stars (or UXORs) are a subgroup of Herbig AeBe stars that exhibit large photometric variability, with irregular dimmings by 2--3 magnitudes in the optical lasting from few days to few weeks \citep{herb99}. These events are accompanied by an increase of the polarization degree in the blue and a rotation of the polarization angle \citep{ab90,natta2000,gri00}. UXORs were found to differ from other AeBe stars by that their circumstellar disks are seen almost edge-on, within about 20 degrees, as evidenced by mm-observations and VLTI interferometry \citep{natta97,kreplin13,kreplin16}. The generally accepted interpretation of UXORs' variability is that it is caused by irregular occultations of the central star inhomogeneities within the disk, such as dust clumps, planetesimals or other disk structures \citep{natta2000,testi01,dull03,gri17}. Thus, when the stellar brightness is significantly reduced during occultations, subtle details of the star and planet formation processes in the vicinity of the star become observable, especially in the blue and UV. These include scattering on small dust grains and hot gas emission from the inner disk. 

\subsection{Herbig Disk Structure}
The spectral energy distributions (SED) of Herbig stars typically have an excess in the infrared and millimeter wavelengths, based on which circumstellar matter, in the form of a disk and sometimes with an envelope, have been inferred. \citet{meeus01} proposed to classify Herbig stars into two groups based on their SEDs: group I whose continuum at the IR to sub-mm wavelengths can be fit as a power law plus a black body, and group II whose continuum can be fit by a power law alone. \citet{dullemond04shadowing} found that group I sources can be modelled as flared disks, while group II sources may be self-shadowed disks with the outer part collapsed in the vertical direction and with direct access to starlight denied. \citet{dullemond04dustsettling} further suggested that the reason for the reduced opacity in the outer disk in group II may be grain growth and settling. Imaging observations later revealed that many disks in group I have a large inner cavity or gap \citep{honda12, canovas13}.

Spatially resolved observations of Herbig stars capable of revealing the extent of and morphology of their circumstellar disks have ongoing for decades (e.g., \citealt{mann97}). Today, these observations are usually carried out in two avenues: scattered light imaging at optical to near-infrared wavelengths that sample outer ($>$15 AU) disk surface layers, and interferometric observations at near-infrared, mid-infrared and (sub-) mm wavelengths. Both dust and gas emission can be observed in the latter. 

In recent years, many structures have been discovered in imaging observations of Herbig disks, including spirals \citep[e.g.,][]{muto12,v21}, gaps \citep[e.g.,][]{perez19hd169142,a18,v19,v21}, large scale asymmetries \citep[e.g.,][]{v13,dong18mwc758,a18}, and shadows \citep[e.g.,][]{avenhaus14}. While the origin of observed structures in many cases are still in debate, an exciting hypothesis supported by more and more evidence is that they are produced by the gravitational interactions between the disk and companions, with some of them being giant planets \citep[e.g.,][]{zhu14stone, dong15spiralarm, facchini18}. Companion-disk interaction models have succeeded in reproducing the general morphology of Herbig disks at multi wavelengths \citep[e.g.,][]{baruteau19,z18}, and can explain the fine details in e.g., the different cavity sizes in different tracers \citep{dong12cavity, zhu12}. In some cases, the companion sculpting the disk has been found and the connection has been confirmed \citep[e.g., HD 100453;][]{dong16hd100453, wagner18, rosotti19}, while in some other cases candidates have been found, but are waiting to be confirmed \citep[e.g., MWC 758;][]{wagner19}. 

Assuming observed structures in Herbig disks are mainly produced by giant planets at a few to a few tens of AU, the inferred occurrence rate of giant planets forming in Herbig disks has been compared with that revealed in direct imaging observations \citep[e.g.,][]{dong16td, dong18spiral}. For certain disk structures, such as spiral arms, it has been found that more giant planets are needed to sculpt the disk than those found in direct imaging surveys, prompting the ``missing planet'' problem and the suggestion that young planets may have low luminosities during most of their lifetime in disks \citep{brittain20}.


\begin{figure}
\begin{center}
\includegraphics[scale=0.5,angle=0]{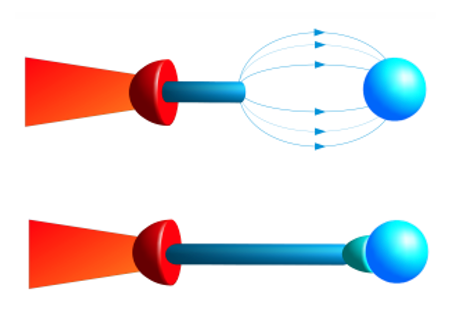}
\caption{Figure adopted from \protect\citet{mendigutia2020}.  The different accretion mechanisms in the presence and absence of a magnetic field respectively. In the top panel, the disk has an inner hole due to the strong magnetic field. Any accreting material then spirals along the field lines and crashes onto the stellar surface, resulting in an accretion shock. The lower panel denotes the situation where there is no magnetic field truncating the disk, which consequently reached the stellar surface. In the Boundary Layer (indicated in cyan), the rotating material rapidly slows down, converting kinematic energy into radiation.} 
\label{accretioncartoon}
\end{center}
\end{figure}

\subsection{Accretion in Herbig Ae/Be Systems}

While solar-mass pre-Main Sequence T Tauri-type stars are widely accepted to be accreting material from their circumstellar disk flowing along magnetic field lines originating in stellar magnetic regions with kG-fields (magnetically controlled accretion, hereafter MA), the situation is not clear for the intermediate mass Herbig Ae/Be systems.  Lower mass, lower temperature objects have convective envelopes in which a magnetic field is generated through the dynamo motion of the charged convective cells.  The field itself truncates the circumstellar disk at a distance of a few stellar radii. From this distance, the accreting disk material free-falls at high velocities onto, and shocks, the stellar photosphere (\citet{Bertout1989}, see Figure \ref{accretioncartoon}). The released gravitational potential energy is re-radiated at UV wavelengths.  The MA paradigm is now widely accepted to act in the case of low mass stars. A review describing much of the evidence is provided by \citet{Bouvier2007}. Eventually, the parental cloud disperses and planets are thought to form in the disk remnant.

\begin{figure*}
\begin{center}
\includegraphics[scale=0.3,angle=0]{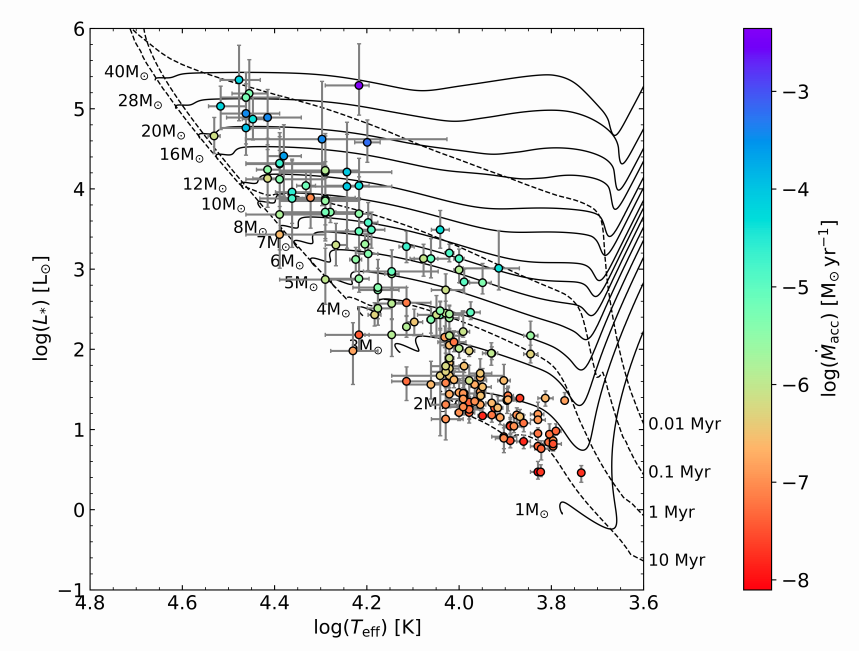}
\includegraphics[scale=0.33,angle=0]{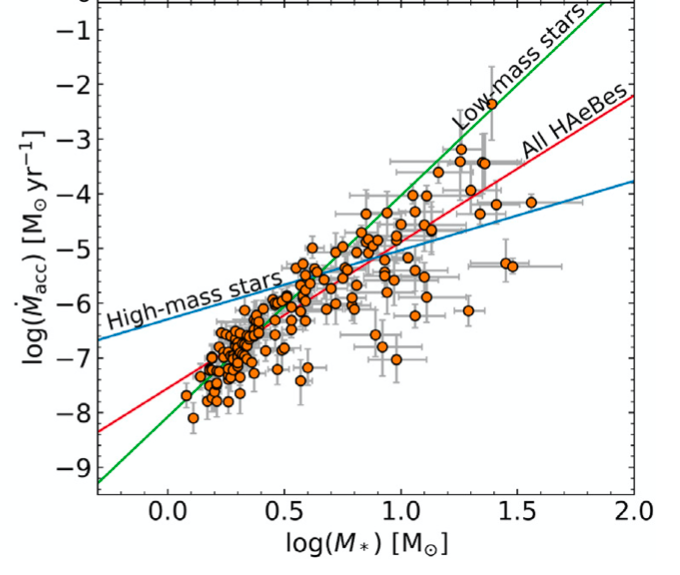}
\caption{Figures from \protect\citet{w2020}. The left panel shows known Herbig Ae/Be stars for which mass accretion rates have been measured. There is a strong correlation in the sense that the higher mass stars have higher accretion rates. The right panel shows that there is a break in the slope of this increase around the 4\msun\ mark. } 
\label{wichfig}
\end{center}
\end{figure*}

The situation for massive stars is still unclear. At present, it is not known how disk accretion operates in massive stars. As these stars are generally non- or weakly-magnetic due to their radiative envelopes, their disk accretion mechanism has to be different to that for the lower mass stars. Indeed, there is growing evidence that for higher mass stars, 
the accretion is different from that in 
lower mass objects.

Remarkably, this transition appears to occur at higher masses than would be expected based on stellar properties alone. The latter would place the transition around the G/F to A spectral boundary where stellar envelopes become radiative and would not be able to generate and sustain a magnetic field. However, a mass of 4 \ \msun, corresponding to mid- to late B spectral type, for this transition was derived by \citet{w2020}. They reported on a clear break in accretion rate properties between T Tauri and Herbig Ae stars on the lower mass side and Herbig Be stars on the higher mass side (Figure \ref{wichfig}). 

\begin{figure*}
\begin{center}
\includegraphics[scale=0.5,angle=0]{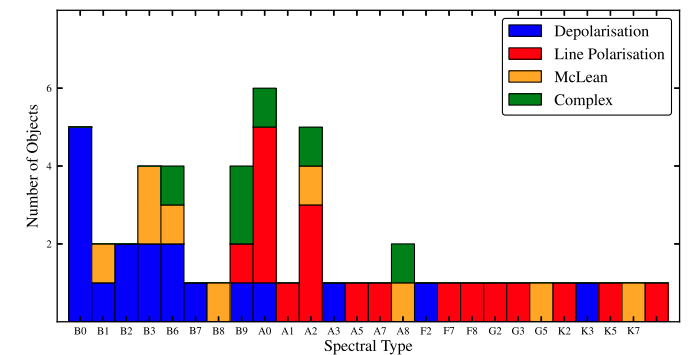}
\caption{Figure adopted from \protect\citet{ababakr2017}. The linear spectropolarimetric properties around the H$\alpha$ line are shown as a function of spectral type, with the detailed effects described in \citet{ababakr2017}.  The observed H$\alpha$ polarization signature appears to change from intrinsic line polarization for spectral sub-types later than B7-B8 to depolarization effects for spectral sub-types earlier than B7-B8.  Although exceptions exist, this general trend was interpreted by \citet{ababakr2017} to indicate a change from magnetospheric accretion (for B7-B8 and later sub-types) to a different manner of accretion (for earlier sub-types).} 
\label{specpol}
\end{center}
\end{figure*}

This higher mass would lead to the surprising conclusion that Herbig Ae stars might be accreting in a similar manner to the magnetic T Tauri stars. It turns out that this is consistent with other findings, much of which is reviewed in \citet{w2020}. Here we draw attention to the linear spectropolarimetric signatures across H$\alpha$ which are observed to be very similar in both T Tauri and Herbig Ae stars. The spectropolarimetric behavior can be explained with accretion hot spots due to magnetically channeled accreting material \citep{Vink2003,ababakr2017}. Magnetic fields were reported to be present in some HAeBe stars (\citep{hubrig2013,alecian2013}), although it is not known yet whether these concern fossil fields or freshly generated magnetic fields.  In contrast, the H$\alpha$ spectropolarimetric signatures found towards Herbig Be stars are more consistent with a disk reaching onto the stellar surface, in much the same way as seen in such data of classical Be stars \citep[e.g.][]{Poeckert1976,Poeckert1977}.  
The study of a large sample of Herbig Ae/Be stars by \citet{ababakr2017} indicates a transition around the B7-8 spectral type (Figure \ref{specpol}), consistent with the 4~\msun \, found by \citet{w2020} based on other diagnostics. 

In summary, there appears to be a break at around late B spectral type where the star formation mechanism changes from magnetically controlled accretion from disks with inner holes to this as yet unknown mechanism. The natural next question is which mechanism takes over from MA at higher masses. A major clue for this may be found in \citet{ababakr2017}'s data which indicate that their disks reach all the way to the stellar surface. Given that, contrary to their circumstellar disks,  stars typically rotate slower than Keplerian, the disk material must lose substantial amounts of angular momentum and therefore kinetic energy before reaching the stellar surface. This situation is very much alike the Boundary Layer (BL) accretion which is known to act in accreting astrophysical systems such as White Dwarfs (e.g. \citealt{Tylenda1977,Romanova2012}). The thin BL region, is where the material reduces its (Keplerian) velocity to the slower rotation of the star when it reaches the stellar surface, and it is here that kinetic energy will be dissipated. The BL has been suggested a number of times to act in Herbig Be stars (e.g. \citealt{Cauley2014}), but has never been adapted and tested for massive objects.

\subsection{Impact of MA vs BL Accretion}

It should be noted that this lack of understanding of the accretion mechanism also directly affects the mass accretion rates that are determined. In case of free-falling material occurring in MA, the accretion luminosity can be roughly approximated to be the gravitational potential energy, $\frac{GM\dot{M}}{R_*}$. However, in Boundary Layer accretion, the energy is released through the slowing down of material from Keplerian rotation to the stellar rotation speed. For very slowly rotating stars, the accretion luminosity approaches half the gravitational potential energy, as essentially all kinetic energy of the orbiting material can be radiated away.  The consequence is that the mass accretion rates are underestimated by a factor of 2 if the MA paradigm is used to determine the rates instead of the BL. The situation is more severe in cases where the star is rotating rapidly. Here, the disk material does not lose as much kinetic energy as in the slowly rotating star case resulting in less energy being radiated away. In addition, the rotational kinetic energy is not only converted into radiation, but part is also used to spin up the star. The  general expression for the luminosity radiated from the Boundary Layer, and in the approximation of a very thin BL, can be written as (e.g. \citealt{lynden1974}):

$$L_{BL} = 
\frac{1}{2} \frac{GM\dot{M}}{R_*} \left(1-\frac{\Omega_*}{\Omega_K}\right)^2
$$
\smallskip

with $\Omega_*$ denoting the angular velocity for the star and $\Omega_K$ for the Keplerian disk material respectively. We thus find that for objects rotating close to the break-up speed the accretion luminosity is much less than in the MA case, and, in turn, the resulting mass accretion rates are underestimated even more. For example, in the case of a star rotating at half the Keplerian speed, the BL accretion luminosity is only 12.5\% of the available gravitational potential energy, and mass accretion rates derived using MA are underestimated by an order of magnitude.

Finally, the inevitable conclusion is that the magnetically controlled accretion scenario turns out to be operating at higher temperatures and masses than previously thought. To further address and investigate this issue, we need dedicated observations tracing the inner disks of a large sample of object.

\subsection{Inner Disk Variability Phenomenon}

Important inner disk regions also include the transition from the ionized accretion zone to “dead zones” beyond the accretion regions where most planets are thought to form or migrate to \citep{mor09,mor09b}. These inner few AU disk regions are inferred to be highly structured from variability studies, but their detailed geometry remains poorly constrained \citep{tann08,sitko08,mor11,k12,men13,reb14,debes17,rod17,kobus20,klu20}. Mis-aligned/warped inner disks and inflated disk walls, inferred from variability studies \citep{wis08,rod17,ben18,rich19,zhu19,klu20,ans20} have been attributed to dynamical sculpting from planets \citep{ben18,zhu19,klu20,ans20}, or stellar \citep{prad86,cat87,bou97}, or disk emission \citep{sitko08}. 

Magnetospheric accretion can also affect the innermost circumstellar structure \citep{ans20,kuff21}. The base of accretion columns in the disk may contain warm optically thick dust, causing a dusty warp in the inner disk. As the magnetic field co-rotates with the star, partial obscurations of the star by such a dusty warp can occur quasi-periodically \citep{Bouvier2007,ans20}. This concept was developed to explain narrow quasi-periodic brightness dips of a T Tau-type star AA Tau, which became a prototype of the dipper phenomenon in low-mass, pre-main-sequence stars with the occurance rate of at least 30\% \citep{rogg21}.  Mapping the time evolution of the geometry of the inner regions of planet-forming disks from the stellar surface to $<$few AU is needed to discriminate between competing mechanisms that could drive observed variability, and help identify geometrical structures that could point to active planet formation in the inner few AU regions of these systems.

\section{Inner Disk Observational Diagnostics}
\label{challenges}

The detailed geometry of the innermost regions around Herbig AeBe disks, from the stellar photosphere to $<$ 5 AU, is poorly understood largely because this circumstellar region is difficult to diagnose with existing instrumentation.  Near-IR extreme-AO facilities such as Subaru's SCExAO/CHARIS \citep{Groff2016} and VLT's SPHERE/IRDIS \citep{sphere2020} are limited to inner working angles of 0$\farcs$1 - 0$\farcs$15 (14-21 AU for a star at 140pc). ALMA can provide exquisite angular resolution imagery of the cold gas and dust content of Herbig AeBe stars, but is not suited to probe warm material close to the host protostar. 

\subsection{Optical/near-IR Interferometry}

Optical and near-IR interferometers have made significant strides in  resolving and/or placing constraints on the emission within the inner 1 AU of Herbig AeBe disks \citep{Per2016,Men2017,varga18,Hone2019,Lab2019,var21,kok21,lopez21}.  Such observations have provided some initial geometrical and physical conditions on the innermost disk regions, such as the inner radius of the dust sublimation zone and information on the vertical scale heights, and dust size distribution and composition (e.g. \citealt{ragland2009,matter2014,matter2016}).  Simplifying assumptions have been made during image reconstruction and geometric model fitting, depending on the quality of the data and the Fourier plane (u,v) sampling, such as was done in early surveys of Herbig stars, e.g., \citealt{millan99,monnier2006}.  

Aperture masking interferometry, which sampled the (u,v) plane very well with the Keck telescope, yielded the first J, H, and K band images of Herbig Ae/Be stars, for example, LkHa 101 \citealt{tuthill2001,tuthill2002} and MWC 349A \citep{danchi01}.  This led to the development of the concept of puffed up inner dust rims  (e.g., see the review by \citealt{dull10}), and motivated imaging with long baseline interferometry at facilities like CHARA and the VLTI.  

More recently the VLTI/GRAVITY K-band survey of 27 Herbig AeBe stars determined the half-flux radius of observed emission in their sample to be 0.1-6 AU \citep{gravity2019}.  The GRAVITY survey also presented some evidence of gaps, some inner disk misalignments, and inferred the potential presence of non-axisymmetric distribution of material from the analysis of their closure phase data \citep{gravity2019}. 

The VLTI GRAVITY (K band) and MATISSE (L, M, and N bands) instruments produce 6 visibilities and 4 closure phases (3 independent) for each spectral channel. Images can be reconstructed particularly for brighter objects if the Auxiliary Telescopes (ATs -- 1.8 m) are used.  These can be reconfigured to sample the (u,v) with baselines as short as about 8 m to as long as 140 m.  Reconstruction of images for fainter objects is much more difficult using the Unit Telescopes (8.4 m) since they are fixed in position. Detailed information the capabilities and performance of the MATISSE instrument is available \citep{lopez21}.

Deriving detailed non-axisymmetric structure from near-IR interferometry is challenging, especially when the timespan of observations are comparable to the dynamical timescale \citep{klu20}. Moreover, depending on the Fourier (u,v) plane coverage, simplifying assumptions are often made during image reconstruction and geometric model fitting \citep{gravity2019,klu20}. Only UV spectro-polarimetry allows observations right down to the stellar surface, the ultimate destination of ISM/stellar interaction. A dedicated survey that used complimentary techniques that are capable of characterizing the inner geometry of Herbig AeBe systems down to the stellar surface, like UV spectropolarimetry, could help to both confirm and better characterize indications of inner disk geometries that are emerging from the IR interferometry community.


\begin{figure*}[!ht]
\begin{center}
\includegraphics[scale=0.1,angle=0]{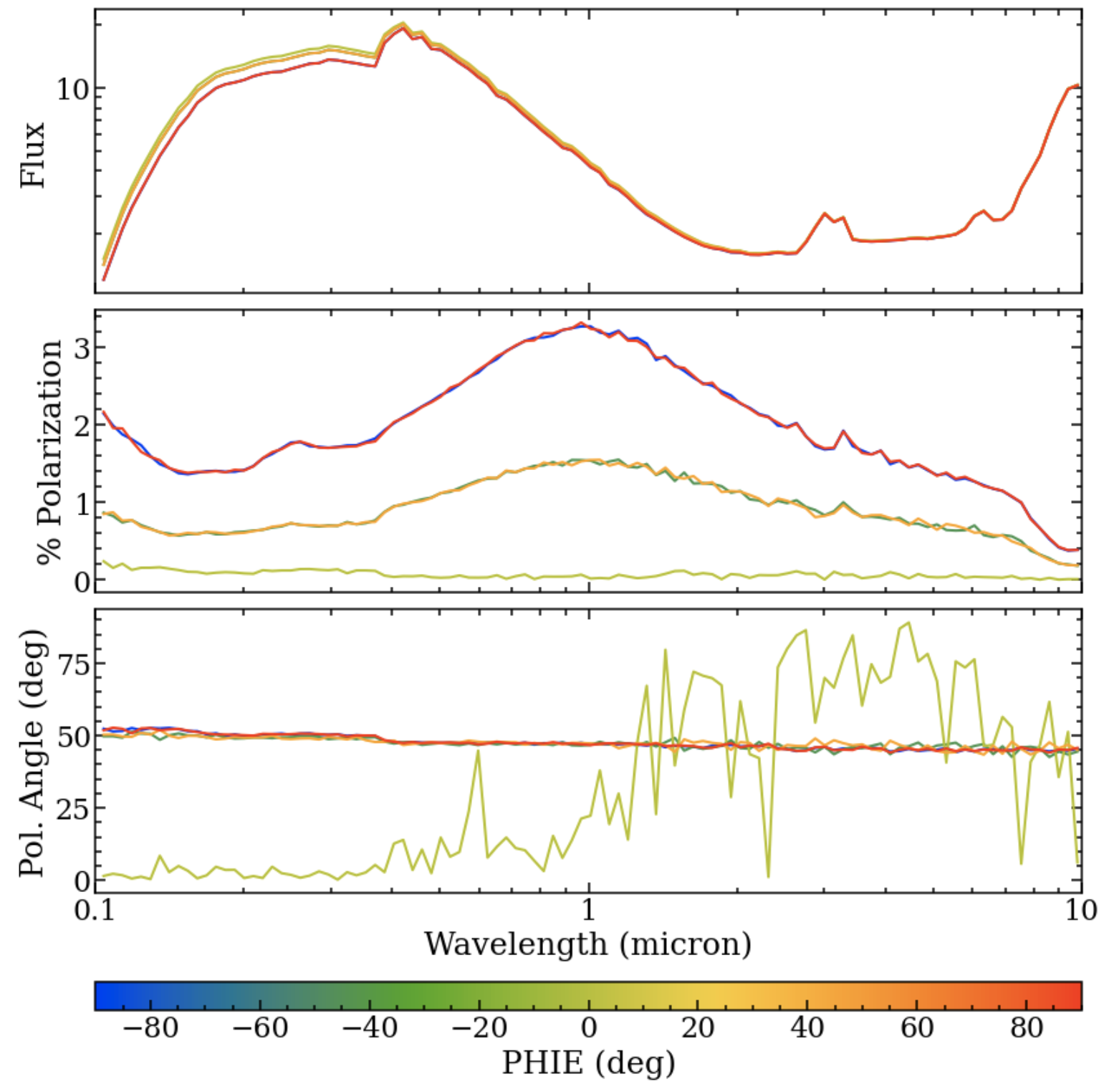}
\includegraphics[scale=0.1,angle=0]{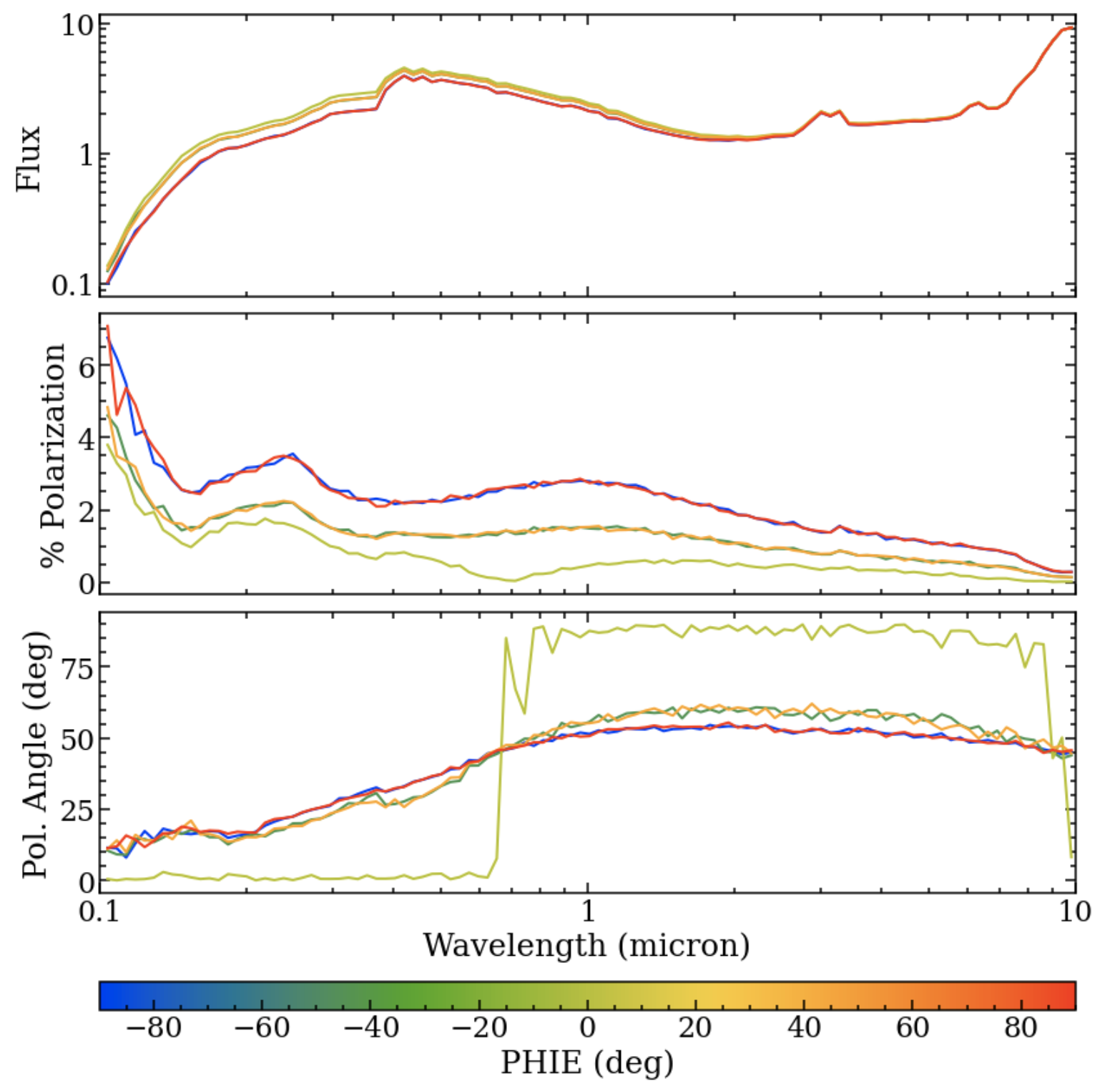}
\includegraphics[scale=0.1,angle=0]{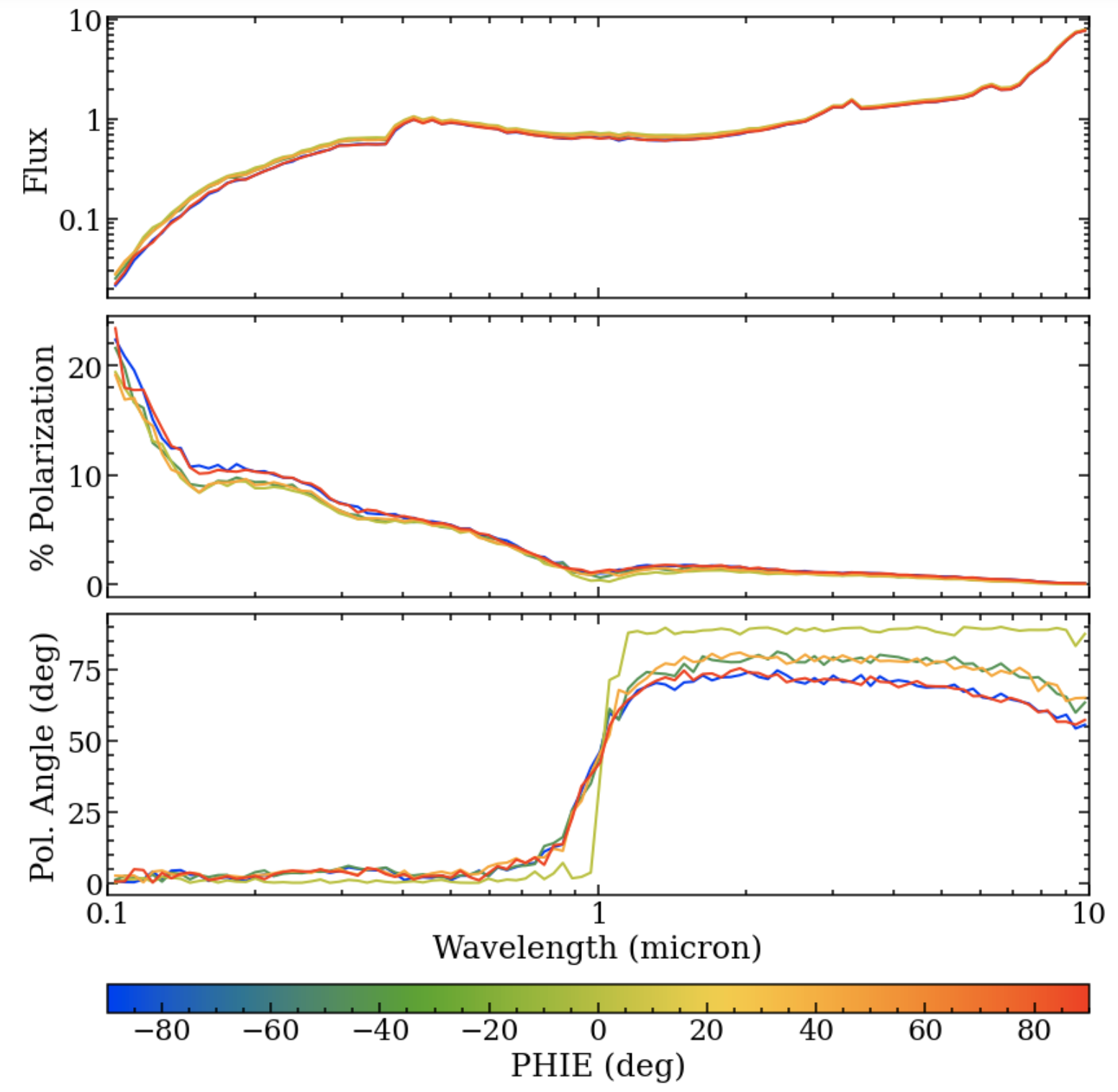}
\caption{HoCHUNK3D MCRT models of a representative Herbig Ae system at 3 different inclinations, ranging from i=12$^{\circ}$ (pole-on), i=42$^{\circ}$ (moderate inclination), to i=72$^{\circ}$ (edge-on).  The distinctly different wavelength-dependent amplitude and time evolution of the polarization and position angle reflect different contributions from the equatorial versus polar regions.} 
\label{inc}
\end{center}
\end{figure*}

\subsection{The Role of UV Spectropolarimetry}\label{roleofuv}

\begin{figure*}[!ht]
\begin{center}
\includegraphics[scale=0.7,angle=0]{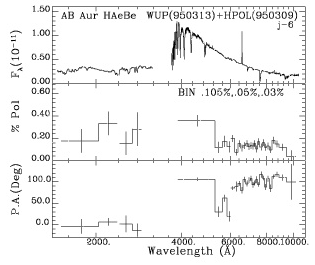}
\includegraphics[scale=0.71,angle=0]{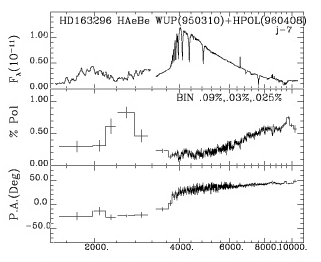}
\caption{Near-contemporaneous optical (HPOL; \citealt{wolff96,davidson14} and UV (WUPPE; \citealt{Ken94}) single-epoch snapshot observations of the 
Herbig Ae systems AB Aur (left) and HD 163296 (right) exhibit distinctive 90 degree position angle flips, indicative of a change from equatorial-dominated to polar-dominated scattering \citep{w2003a,w2003b,w2004}. Figure courtesy of Marilyn Meade.} 
\label{wuppe}
\end{center}
\end{figure*}

Linear spectropolarimetry can help disentangle the geometry of otherwise unresolved circumstellar environments (see e.g. \citealt{bjo00}). For pre-main sequence stars, UV spectropolarimetry provides a unique diagnostic of the innermost polar regions of protoplanetary disks, including the presence and location of accretion hot spots, inner warps and misaligned disks, and inflated disk rims \citep{whitney13}. Critically, the dominant opacity source of young protoplanetary disks changes from equatorial (disk) scattering to polar scattering as one moves from the optical to UV, leading to a distinctive 90 degree flip in polarization position angle (PA)\citep{w2003a,w2003b,w2004}. 

One way to visualize the utility of this diagnostic is to construct HoCHUNK3D MCRT models for a representative Herbig Ae star \citep{whitney13}, building upon diagnostics already presented in \citet{w2003a} and \citet{w2003b}. Here we use a model constructed for the Herbig Ae star HD 163296, used to match optical/IR/mm photometric and optical coronagraphic imagery as detailed in \citet{rich19}.  We have increased the density of polar material as compared to that used by \citet{rich19}, to produce a 90$^{\circ}$ polarization PA flip as observed in the system by WUPPE (Figure \ref{wuppe}); i.e. constructing a MCRT model for the system in the absence of UV spectropolarimetry led to a severe underestimation of the amount of polar material in the system. As shown in Figure \ref{inc}, we have allowed the system inclination of this benchmark simulation to vary to preferentially sample more of the equatorial (inclination = 72$^{\circ}$) and polar regions (inclination = 12$^{\circ}$), and show a range of system viewing angles (PHIE angles). 
For moderate (i=42$^{\circ}$) to more equatorial (i=72$^{\circ}$) inclinations, the predicted polarization in the UV dramatically increases to reach p = 5-20\%.  Moreover, these simulations demonstrate how the changing opacity from being equatorially dominated to polar dominated can produce distinctive changes at many system inclination angles. For example, in more edge-on inclination systems (i=72$^{\circ}$), a distinctive 90$^{\circ}$ flip in the wavelength-dependence polarization PA is clearly seen as one moves towards bluer wavelengths, that depict where scattering from the polar region becomes more dominant than equatorial scattering.  Note the specific wavelength at which this flip occurs will depend on equatorial and polar dust masses and dust properties assumed.  A more gradual change in polarization PA is evident at moderate inclinations (i=42$^{\circ}$).  By contrast, when viewing a system nearly pole-on (i=12$^{\circ}$), both the amplitude of the wavelength-dependent polarization decreases and the presence of a PA flip disappears, as one is exclusively dominated by scattering from the polar regions.

\subsection{WUPPE and HST-FOS}

A limited number of UV spectropolarimetric observations of Herbig AeBe stars have been made. The Wisconsin Ultraviolet Photo-Polarimeter Experiment (WUPPE) flew on the ASTRO-1 and ASTRO-2 space shuttle missions. WUPPE obtained single-epoch observations of 4 of the brightest Herbig AeBe stars \citep{rsl92,bjo94,bjo95,bjo96,bjo00}. Interestingly, the distinctive 90$^{\circ}$ polarization PA flip noted in Subsection \ref{roleofuv} was seen in 3 sources, including HD 45677 \citep{rsl92}, HD 163296 (Figure \ref{wuppe}; see also \citealt{bjo95}), and AB Aur (Figure \ref{wuppe}). The single epoch observation of HD 50138 suggested that its inner polar region was more deficient of dust \citep{bjo98}, and questions remain about whether the system is a pre-main sequence or post-main sequence system.  The short durations of both WUPPE flights precluded the type of time series UV spectropolarimetric survey that we envision for Polstar.

The Hubble Space Telescope FOS also enabled UV spectropolarimetry. However, because of the 1-$\sigma$ accuracy of the linear polarization of the instrument pre-COSTAR (0.3\%) and post-COSTAR (0.5\%) (see HST-FOS ISR 150), use of HST-FOS UV spectropolarimetry for Herbig AeBe science, including time series UV spectropolarimetry, was of more limited utility while the instrument was on HST.

\subsection{Polstar}
Polstar is a 60cm high resolution UV spectropolarimetry MIDEX mission whose design and capabilities are presented in \citet{sco21} and \citet{sco22}. Polstar can operate in two different observing channels: channel 1 provides R$>$30,000 spectropolarimetry from 122-200 nm in all four Stokes parameters (I,Q,U,V), and channel 2 provides R$>$140 spectropolarimetry from 122-320nm in all four Stokes parameters (I,Q,U,V), with additional pure spectroscopic coverage extending out to 1000nm. The mission is designed to provide 0.01\% precision in all four Stokes parameters during its 3 year baseline mission, 30x-50x better than that achieved by HST/FOS. The prospective science investigations we detail below are based on simulated performance specifications for Polstar based on the concept outlined in \citet{sco21}. Additional key science cases for Polstar are presented in \citet{andersson21}, \citet{gayley21}, \citet{jones21}, \citet{peters21}, \citet{shultz21}, and \citet{stlouis21}.

\section{Potential Time Domain UV Spectropolarimetric Surveys}
\label{observations}

\begin{figure*}[!ht]
    \includegraphics[width=\textwidth]{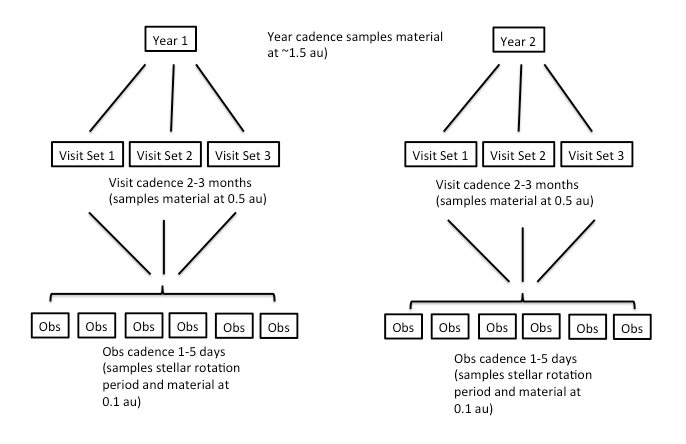}
    \caption{Our suggested Polstar nested cadence structure
    \label{fig:cad}
    }
    \end{figure*}

The expected performance capabilities of Polstar \citep{sco21} could enable time domain UV spectropolarimetric investigations. In particular, we identify two surveys that could 1) Test the hypothesis that magneto-accretion operating in young planet-forming disks around lower-mass stars transitions to boundary layer accretion in planet-forming disks around higher mass stars; and 2) Discriminate whether transient events in the innermost regions of planet-forming disks of intermediate mass stars are caused by inner disk mis-alignments or from stellar or disk emissions. We detail these two representative surveys below.

\subsection{Accretion Diagnostics} \label{accretiondiagnostics}
\subsubsection{Survey Design}
Polstar's expected performance capabilities \citep{sco21} could enable a survey to test the hypothesis that magneto-accretion operating in young planet-forming disks around lower-mass stars transitions to boundary layer accretion in planet-forming disks around higher mass stars. We have identified a representative sample of 30 bright, nearby Herbig stars that equally sample the range of spectral sub-types, i.e. 10 Herbig Be stars, 10 early-type Herbig Ae stars, and 10 late-type Herbig Ae stars (see Table \ref{tab:x1targets}). This representative sample is purposefully comprised of nearby, bright, ``well studied'' Herbigs. In particular, we note that these systems have good distances from GAIA \citep{dr2}, stellar masses and luminosities, age estimates, extinction, system inclinations, and known IR excesses \citep{blondel2006,w2020}. Many have the surface regions of their outer disks spatially resolved with optical/NIR coronagraphic observations, and their outer dust and gas distributions resolved by ALMA. Owing to this wealth of data, baseline models of the outer regions of many of these disk systems already exist in the literature.

We suggest a survey that utilizes a series of nested visits using Polstar's Channel 2. The use of Channel 2 will provide R$>$140 linear spectropolarimetry from 122-320 nm as well as spectra from 122-1000 nm. We computed integration times using the latest version of Polstar's exposure time calculator, and provide the per-visit total integration time per waveplate position as well as the total mission exposure time anticipated (Table \ref{tab:x1targets}).  Our exposure times have been computed to provide 0.1\% precision linear spectropolarimetry, per resolution element or after binning to 5 nm-size bins, per visit. 

We propose the nested visit cadence shown in Figure \ref{fig:cad}, which probes a number of critical time-scales. First we recommend a series of 6 visit sets, with each visit set occurring at a cadence of 1-5 days, to sample the stellar rotation period of each target. Because our brightest sources utilize individual exposure times approaching 1-2 seconds per waveplate position, we plan to stack several short repeat observations to build up sufficient SNR, which will provide a limited amount of ultra-short (minute) variability information.  Next, we aim to repeat the above stellar rotation sequencing in 3 visit sets separated by 2-3 months each, to probe orbital rotation at 0.1 AU in each system. Our final nested loop is to repeat the entire aforementioned process in year 2 of the mission, to probe orbital rotation at $\sim$1 AU in each system.  In total, this sequence of nested visits should yield $\sim$36 unique epochs of observations per target.

\subsubsection{Analysis}

\begin{figure*}[!ht]
\begin{center}
\includegraphics[scale=0.4,angle=0]{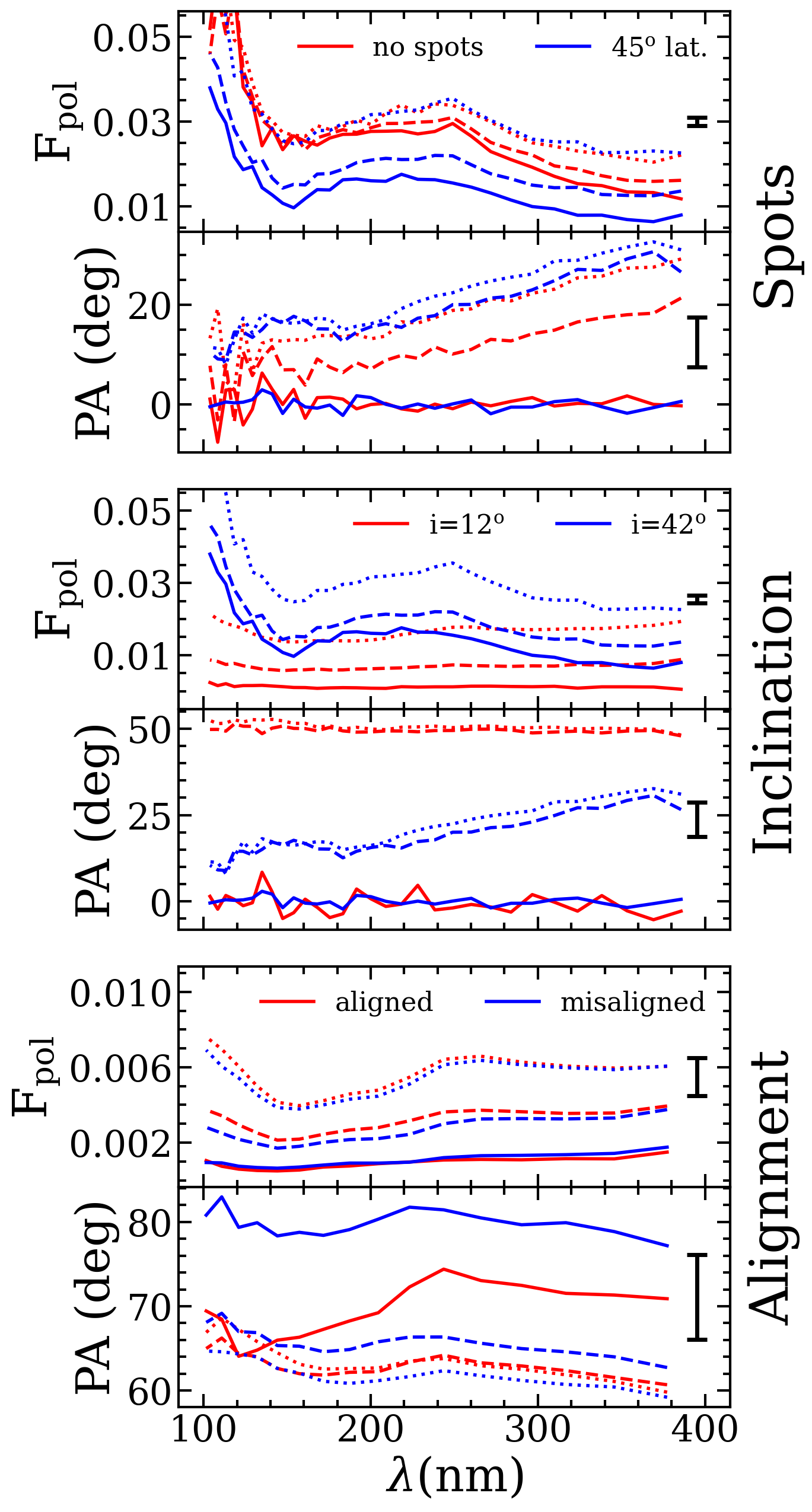}
\caption{The amplitude and wavelength dependence of UV fractional polarization and Position Angle (PA) exhibit distinctive trends for different accretion hot spot models, system geometries (inclinations), and aligned/ mis-aligned inner disks, including different evolution over time (depicted by solid, dashed, dotted lines). The amplitudes of these trends are easily discernible given the expected 1e-3 polarization precision Polstar will achieve for each target based on our exposure times and/or binning.} 
\label{kellen}
\end{center}
\end{figure*}

Magnetospheric accretion (MA) is characterized by a truncated inner disk with accretion funnels bringing material to hot spots at high latitude, whereas the emitting region of boundary layer (BL) accretion should extend to the star in an equatorial annulus (Figure \ref{accretioncartoon}). MA will therefore exhibit different time-dependent geometry in the polar direction as compared to BL accretion. Because UV spectropolarimetry is able to discern geometrical structure down to stellar surface, a dedicated UV spectropolarimetric mission like Polstar could determine the transition between MA and BL accretion. 

The time and wavelength dependence of the continuum polarization and polarization PA of the dedicated survey described in Section \ref{accretiondiagnostics} would be modeled using 3D Monte Carlo Radiative Transfer (MCRT) radiative transfer codes such as HoCHUNK3D \citep{whitney13} and RADMC-3D \citep{rad}. HoCHUNK3D allows for multiple component equatorial and polar scattering regions and accretion hot spots; it can predict full SEDs and integrated (unresolved) polarization as a function of wavelength for a range of model viewing angles. One could use the latter as a proxy for comparison to the relative time dependence of continuum polarization and PA changes.

Using the known properties of our well-studied targets compiled in Table \ref{tab:x1targets} (e.g., inclination, star properties) to minimize degeneracies, one could compute model grids to sample a range of inner disk radii and accretion hot spot number and latitude to compare against predictions for MA vs. BL accretion. The predicted UV continuum fractional polarization and PAs for accretion-spot free systems (BL) exhibit different amplitude and time-dependent behavior than MA systems with moderate or high latitude hot spots (Figure \ref{kellen}), at levels that significantly exceeds the expected 1e-3 fractional polarization precision that Polstar will achieve for each of our targets. This predicted polarization behavior also significantly exceeds the Polstar mission linear polarization precision floor of 1e-4.

Figure \ref{kellen} details one aspect of the differences expected between the time-dependent, wavelength-dependent linear spectropolarimetric signatures expected from MA (hotspots) vs BL accretion. Similarly, inclusion of different inner disk geometries expected for MA vs BL accretion (Figure \ref{accretioncartoon}) will yield different time-dependent signatures. Although not a mission performance requirement, should Polstar extract useful polarimetric signal beyond 320nm, this will be used to constrain the expected PA flip signifying a change in polar vs equatorial scattering.  

Superimposed on the intrinsic polarization signal that we aim to analyze will be interstellar polarization (ISP), that exhibits its own distinctive wavelength-dependent, time-independent signature \citep{serk}. The amplitude of ISP towards our sources is expected to be low and indeed our analysis could proceed in the absence of removing ISP.  Nevertheless, we will use standard techniques including field star polarization, de-polarization across emission lines, and the time dependence of the observed total polarization to characterize and remove the ISP component (see e.g. \citealt{w2003,w2006,w2007,nord01}). Since the Polstar mission includes a significant exploration of interstellar polarization (e.g. \citealt{sco21} and Andersson et al 2021), our ISP characterization and removal efforts will be aided by our broader team's expertise in this science.

The anticipated broad UV-optical spectrophotometric capabilities of Polstar will also enable us to constrain MA vs BL accretion via analysis of each spectra we obtain. We will measure the excess emission present in our de-reddened spectra across both the Balmer jump and in $\sim$100 \AA\ binned UV continuum regions. The ratio between UV and Balmer Jump excesses is predicted to be different for BL vs MA models for a given star (e.g., Figures 4-5 in \citealt{mendigutia2020}). We anticipate measuring excesses ranging from 20-150 milli-mag (e.g. \citealt{fairlamb, mendigutia2020}, which is well within the projected 5 milli-mag broad photometric precision anticipated from Polstar. Polstar’s spectrophotometric capabilities and observing cadence will allow us to simultaneously measure the UV excess across the continuum, accretion-sensitive lines (Mg II, CIV), and the BJ (P1) to test BL vs MA predictions, as described in \citet{mendigutia2020}.

\subsection{Inner disk geometry}
\label{sec:innerdiskpol}
\subsubsection{Survey Design}

Polstar's expected performance capabilities \citep{sco21} could also discriminate whether transient events in the innermost regions of planet-forming disks of intermediate mass stars are caused by inner disk mis-alignments or from stellar or disk emissions.  We propose obtaining $\sim$36 epochs of observations of the same 30 Herbig Ae stars from our representative target list (Table \ref{tab:x1targets}) with both Polstar's channel 1 and channel 2. The integration times we computed using the Polstar (Table \ref{tab:x1targets}) will achieve a minimum SNR of $>$25 per resolution element near 150 nm, and often will achieve SNR closer $>$40-50. Following every Channel 1 observation, we will obtain a contemporaneous Channel 2 observation, to provide 
R$\sim$200 linear spectropolarimetry from 122-320 nm as well as spectra from 122-1060 nm. Our channel 2 integration times were computed using the Polstar ETC per waveplate position (Table \ref{tab:x1targets}), using a requirement that we achieve 0.1\% precision linear spectropolarimetry, either per resolution element or per $<$5nm binned element depending on source brightness. We would use the same nested cadence as noted in Section \ref{accretiondiagnostics}, that samples weekly (stellar rotation period and material at 0.1 AU), 2-3 month (material at 0.5 AU), and 1 year (material at ~1.5 AU) times-scales (Figure \ref{fig:cad}).

\subsubsection{Analysis}

Mis-aligned inner disks produce periodic wavelength-dependent polarization and PA variability (see e.g. Figure \ref{kellen}), inner disk wall inflation would produce short time-period periodic behavior, and disk wind variability could produce stochastic variability on the timescale of our observations. Collectively, our time series data will discriminate between competing models for transient events in the inner ($<$few AU) regions of Herbig AeBe disks: inner disk mis-alignments and stellar (or disk) emission.

We will quantify the amplitude, time dependence (periodic, quasi-periodic, aperiodic), and wavelength dependence of variable continuum polarization, continuum polarization PA, and UV-optical continuum flux for our channel 2 observations of each target. Similarly, we will quantify variability in line profiles, velocities, and line strengths of wind lines (e.g., NV 124nm, SiIV 140nm, CIV 155nm, FeIII 190nm) from our channel 1 spectra. Observed variability timescales will guide the location of circumstellar region driving variability, especially as our survey cadence samples orbital timescales of material at 0.1-1.5 AU. The cadence and resolution of our channel 1 observations allow mapping of the type of stellar wind/chromospheric variability observed by IUE for AB Aur (150 km/s modulations on timescales of 40-50 hours; \citealt{cat87}) across our entire target list.

Determining the detailed geometry of the inner circumstellar environments is key to our objective. We will model the time-dependent polarization and PA using detailed 3D MCRT models (HoCHUNK3D, \citealt{whitney13} and/or RADMC-3D \citealt{dull10}). Starting with baseline models for our well-studied canonical targets, our initial model parameter space will include mis-aligned inner disks, inflated inner disk walls, and disk winds. The wavelength-dependent polariation and position angle as a function of time of a mis-aligned inner disk, for example, is shown in Figure \ref{kellen}.  We note that, if needed, HoCHUNK3D and RADMC-3D can accommodate custom density distributions informed by our observations.

\subsection{Enabled Science}

Polstar could enable a wealth of additional science investigations of UV-bright protoplanetary disks. 

Small dust grains, i.e., significantly smaller than the wavelength of the scattered light, can be detected in disks by measuring continuum polarization in the UV. Since disk wind can drag only small particles, the role of the disk wind in influencing the ``dipper phenomenon'' \citep{rod17} can be verified by observing polarization during and outside photometric dips. In particular, objects with confirmed misalignment of the outer and inner disks would be of interest. By comparing UV and optical polarization, a conclusion about the dust composition will be made.

A dedicated or triggered studied of bright UX Ori stars in Polstar's Channel 2 could also hold significant scientific merit. Detecting a large polarization degree ($>10$\%) in the UV continuum during the brightness minima of UXORs would help demonstrate that small grains are the main constituents of the dust population in planet-forming disks of intermediate-mass stars. Moreover, the continuum polarization may also be high in the far UV during the bright state of UXORs. In particular, if small particles dominate the dust population, an increase of polarization towards 150\,nm should be observed, because the contribution of the unpolarized radiation from an A--F star decreases in the far UV, while the polarization resulting from Rayleigh scattering increases. This effect can also be searched for in Ae stars without occultation events. Finally, detecting gas emission lines during UXORs occultation events will characterize the gas composition in their inner disks. 

Such a survey of bright states of UXORs could be complemented by UXOR observations as targets of opportunity during stellar occultation events. Since such events are irregular and cannot be predicted, selected targets could be monitored with ground-based remotely controlled telescopes ($\le 1$\,m) equipped with the optical three-channel DIPOL-2 polarimeter \citep{p2014}. Several copies of DIPOL-2 have been and are being deployed on Hawaii (Haleakala Observatory), the Canary Islands (Tenerife) and Tasmania (Greenhill Observatory). These telescopes will be united in a remote/robotic optical polarimetery network capable of performing continuous monitoring campaigns. Once an occultation event is detected in the optical, simultaneous observations with Polstar in the UV could be triggered. A joint analysis of the optical, near-UV and far-UV polarization could fully characterize the dust population of these young planet-forming systems. Simultaneous optical and UV polarimetric observations of dippers will also help to distinguish between the disk wind (resulting in formation of dust clumps above the plane of the outer disk) and dusty warps in a misaligned inner disk.  

\subsection{Beyond Polstar}

The type of time domain UV spectropolarimetry enabled by a MIDEX mission like Polstar could be expanded to include a wealth of lower-mass, less UV bright T Tauri stars using POLLUX, a proposed high-resolution UV spectropolarimeter for LUVOIR \citep{pol1,pol2}. The detailed accretion properties of 4 T Tauri stars are being diagnosed with time domain UV spectroscopic observations being made by the HST ULLYSES program \citep{hst}. While such stars are too UV faint to study with Polstar, a facility like POLLUX could diagnose the detailed geometry of MA in lower mass stars, extending from the stellar surface out to several AU, in a complementary way that Polstar would do this for Herbig AeBe stars. Similarly, POLLUX would enabled the inner disk geometry of YSO ``dippers'' \citep{rod17} to be explored.

\section{Summary and Anticipated Outcomes}
\label{outcomes}

The Polstar MIDEX mission would unlock a new era in time domain astrophysics, UV spectropolarimetry.  The linear and circular spectropolarimetric capabilities of Polstar would enable a rich breadth of studies across astrophysics, ranging from massive star science, to the interstellar medium, to planet forming young protoplanetary disks.  A targeted time domain survey of 30 Herbig AeBe protoplanetary disks with Polstar, that samples the geometry of material at 0.1-1.5 AU via 36 well-timed epochs of observations, would enable us to

\begin{enumerate}
    
\item Test the hypothesis that magneto-accretion operating in young planet-forming disks around lower-mass stars transitions to boundary layer accretion in planet-forming disks around higher mass stars; and 

\item Discriminate whether transient events in the innermost regions of planet-forming disks of intermediate mass stars are caused by inner disk mis-alignments or from stellar or disk emissions.
\end{enumerate}

We would use the change in the time average (inclination corrected) polar geometry, presence/absence of accretion hot spots, inferred inner gas disk radius, and UV/BJ excess as a function of spectral type to quantify the change from MA to BL accretion, addressing key science question \#1.

We would use changes in wind line profiles and changes in the wavelength-dependent polarization and polarization position angle over time, along with detailed 3D-MCRT modeling, to reconstruct the inner disk geometry for each system. This would enable us to determine the prevalence of mis-aligned inner disks and stellar/disk wind induced phenomenon, addressing key science question \#2.

\acknowledgments
JW acknowledges support from HST-GO 15437 to develop portions of the models presented in this work. PS acknowledges his financial support by the NASA Goddard Space Flight Center to formulate the mission proposal for Polstar.  RI acknowledges funding support from a grant by the National Science Foundation (NSF), AST-2009412. SB acknowledges support from the ERC Advanced Grant HotMol, ERC-2011-AdG-291659.  This research has made use of the SIMBAD database, operated at CDS, Strasbourg, France. This research has made use of NASA's Astrophysics Data System Bibliographic Services.


\begin{deluxetable*}{ccccccc}
    \tablecaption{Representative Target List}
    \tablehead{
    \colhead{Name} & \colhead{Alt Name} & 
    \colhead{Spec Type} & \colhead{Vmag} & 
    \colhead{Channel} & \colhead{Exp Time(s)} & 
    \colhead{Total Duration(s)}
    }

\startdata
HD 53367 & MWC 166 & B0 & 6.96	& Ch 2	& 45 &	28800	\\
HD 53367 & MWC 166	& B0 & 6.96 & Ch 1 & 150	& 39600	\\				
HD 174571 &	MWC 610	& B2 & 8.892 & Ch 2 & 1500 &	378000	\\
HD 52721 &	GU Cma	& B2 & 6.59	& Ch 1 & 60	& 	14400	\\
HD 52721 &	GU Cma	& B2 & 6.59 & Ch 2 & 6	&	14400 \\	
HD 200775 &	MWC 361	& B3 & 7.43	& Ch 1 & 300	&	75600 \\
HD 200775 &	MWC 361	& B3 &	7.43 & Ch 2 & 45	&	28800	\\
HD 259431 &	MWC 147	& B6 &	8.7	& Ch 2 &	120	&	90000	\\
HD 85567 &	V596 Car & B7 & 8.57 & Ch 2 &	120	&	90000	\\
HD 87643 &	MWC 198	& B8 &	9.5	& Ch 2 &	1500 &		378000	\\
HD 150193 &	MWC 863	& B9 &	8.79 & Ch 2	& 1500 & 	378000	\\
HD 37806 &	MWC 120	& B9 &	7.9	& Ch 2 &	120	&	90000	\\
HD 37806 &	MWC 120	& B9 &	7.9	& Ch 1 &	600	&	151200	\\
HD 50138 &	MWC 158	& B9 &	6.67 & Ch 2 &	2 &	14400	\\
HD 50138 &	MWC 158	& B9 & 	6.67 & Ch 1 & 150	& 	39600	\\
HD 250550 &	MWC 789	& B9 &	9.59 & Ch 2 &	500	&	360000	\\
HD 58647 & 	SAO 152860	& B9 & 6.85  & Ch 2 & 20 & 14400	\\
HD 97048 &	CU Cha	& A0 &	9.0 & Ch 2 &	900	& 360000	\\
HD 31293 &	AB Aur	& A0 &	7.05 & Ch 2 &	2	& 	14400	\\
HD 31293 &	AB Aur	& A0 & 7.05	& Ch 1 &	60	&	14400 \\
HD 100546 &	PDS 340	& A0 &	6.3	& Ch 2 &	6	&	14400	\\
HD 100546 &	PDS 340	& A0 &	6.3	& Ch 1 &	20	&	10800	\\
HD 104237 &	DX Cha	& A0 &	6.6	& Ch 2 &	3	&	14400	\\
HD 141569 &	SAO140789 & A0 & 7.12 & Ch 2 &	3	&	14400	\\
HD 141569 &	SAO140789 &	A0 & 7.12 & Ch 1 &	120	& 	28800	\\
HD 179218 &	MWC 614	& A0 &	7.39 & Ch 2 &	120	&	90000	\\
HD 179218 &	MWC 614	& A0 &	7.39 & Ch 1 &	600	& 	151200	\\
HD 190073 &	MWC 325	& A0 &	7.73 & Ch 2 & 	240	& 	180000	\\
HD 163296 &	MWC 275	& A1 &	6.85 & Ch 2 &	2	&	14400	\\
HD 163296 &	MWC 275	& A1 &	6.85 & Ch 1 & 	600	& 	151200	\\
HD 31648 &	MWC 480	& A5 &	7.62 & Ch 2 & 240	&	180000	\\
HD 145718 &	PDS 80	& A5 &	8.8	& Ch 2 &	1800	&	453600	\\
HD 203024 &	SAO 19283	& A5 &	8.8	& Ch 2 &	500	&	360000	\\
HD 100453 &	SAO 239162	& A9 &	7.79 & Ch 2 &	240	&	180000	\\
HD 36112 &	MWC 758	& A8 &	8.27	& Ch 2 &	500	&	360000	\\
HD 139614 &	SAO 226057	& A8 & 	8.24 & Ch 2 &	500	&	360000	\\
HD 169142 &	MWC 925	& A9 &	8.16	& Ch 2 &	240	&	180000	\\
HD 144432 &	SAO 184124	& F0 &	8.19	& Ch 2 &	45	&	28800	\\
HD 142666 &	SAO 183956 & A9 &	8.82	& Ch 2 &	1800	&	453600	\\
HD 199143 & SAO 163989	& F8 &	7.32	& Ch 2 &	240	&	18000	\\
\enddata
\tablecomments{Note the exposure time is the anticipated integration time at each waveplate position at each visit, in units of seconds.  The total duration is the net integration time expected for the target inclusive of all visits, in units of seconds. \label{tab:x1targets}}
\end{deluxetable*}


\clearpage
\newpage

\bibliographystyle{aasjournal}
\bibliography{references}

\end{document}